\documentclass[twocolumn,showpacs,preprintnumbers,amsmath,amssymb]{revtex4}

\usepackage{graphicx}
\usepackage{dcolumn}
\usepackage{bm}

\newcommand{\Vtip}{V_{\textrm{tip}}}

\newcommand{\Vcpd}{V_{\textrm{cpd}}}
\newcommand{\Vtpg}{V_{\textrm{tpg}}}
\newcommand{\Vfb}{V_{\textrm{tpg}}^\textrm{fb}}

\newcommand{\Idot}{I_{\textrm{dot}}}

\newcommand{\Phiin}{\Phi^{\textrm{in}}}
\newcommand{\Phid}{\Phi^{\textrm{d}}}

\newcommand{\eq}{\begin{equation}}
\newcommand{\qe}{\end{equation}}

\newcommand{\didv}{\partial \Idot / \partial \Vtpg}
\newcommand{\atpg}{\alpha_{\textrm{tpg}}}
\newcommand{\atip}{\alpha_{\textrm{tip}}}

\begin{document}

\title{Measurement of the Tip-Induced Potential in Scanning Gate Experiments}

\author{A. E. Gildemeister}
\email{gildemeister@phys.ethz.ch}
\author{T. Ihn}
\author{M. Sigrist}
\author{K. Ensslin}

\affiliation{Laboratory of Solid State Physics, ETH Z{\"u}rich,
CH-8093 Z{\"u}rich, Switzerland}

\author{D. C. Driscoll}
\author{A. C. Gossard}

\affiliation{Materials Department, University of California,
Santa Barbara, CA-93106, USA}

\date{February 22, 2007}

\begin{abstract}

We present a detailed experimental study on the electrostatic interaction between a quantum dot and the metallic tip of a scanning force microscope. Our method allowed us to quantitatively map the tip-induced potential and to determine the spatial dependence of the tip's lever arm with high resolution. We find that two parts of the tip-induced potential can be distinguished, one that depends on the voltage applied to the tip and one that is independent of this voltage. The first part is due to the metallic tip while we interpret the second part as the effect of a charged dielectric particle on the tip.
In the measurements of the lever arm we find fine structure that depends on which quantum state we study.
The results are discussed in view of scanning gate experiments where the tip is used as a movable gate to study nanostructures.
\end{abstract}


\maketitle

\section{Introduction}

Common to most experimental studies on transport in quantum dots is that they investigate the various aspects based on macroscopic current and voltage measurements, not based on local measurements. An interesting goal for a local study of quantum dots would be, for example, to measure the spatial variation of the probability density of the electrons in the dot. A promising approach to this and other questions pertaining to local properties of quantum dots is the scanning gate technique, where the sharp conducting tip of  a scanning force microscope (SFM) is employed as a movable gate that can be scanned over the surface of the sample. The technique has already been successfully used to manipulate single electrons in quantum dots in carbon nanotubes \cite{Woodside:2002} and Ga[Al]As \cite{Pioda:2004,Kicin:2005}, where a singly occupied quantum dot could be studied \cite{Fallahi:2005}. It was also discussed from a theoretical point of view in the context of probability density mapping \cite{Mendoza:2003,Mendoza:2005}. Other nanostructures like quantum point contacts \cite{Topinka:2001} and Aharonov-Bohm rings \cite{Hackens:2006} have likewise been studied.   

In spite of the high number of studies that employ the scanning gate technique, relatively few data are available about an important factor common to all of the experiments, namely the potential that the tip induces in the sample, here called ``tip potential" for brevity. Early on the role of the tip potential for the interpretation of scanning gate data has been mentioned \cite{Finkelstein:2000} but only recently attempts have been made to experimentally determine it \cite{Kicin:2005,Pioda:2007,Gildemeister2:2007}. In previous work it was the quantum system which was studied with the help of the tip. However, implicitly, some of the measurements have revealed just as much information about the potential that the tip induces in the sample \cite{Pioda:2004,Woodside:2002}.

Here we deliberately used a quantum dot as a very sensitive potentiometer to study the tip potential. We demonstrate how, with the help of a feedback mechanism, one can map the tip potential with high spatial and energetical resolution. Additionally, we show how the tip's lever arm on the quantum dot can be mapped and used to better understand the properties of the tip potential. For the measurement of the lever arm we used a technique that minimizes the perturbation of the energy levels of the quantum dot. In these measurements we find fine structure which illustrates how the scanning gate technique may yet unveil information about the quantum dot.

A tip potential useful for probability density mapping would have to fulfill the following criteria: It should be geometrically simple, so as not to complicate the interpretation. The spatial extent should be small compared to the quantum dot or, even better, the Fermi wavelength. The magnitude of the potential should be small compared to the charging energy of the dot or, better, the single level spacing. If these requirements are fulfilled then the tip can be regarded as a small perturbation. In this case the shift in energy of a quantum state due to the tip potential would be proportional to the local probability density of the state. To our knowledge in all published work tip potentials were used that would not be able to fulfill all of these criteria. We will discuss the tip potential in view of these requirements.

\section{Experimental Setup}
In most scanning gate measurements of quantum dots the tip was scanned over the dot at constant height and the current through the dot was recorded as a function of the tip position. If the dot is in the Coulomb blockade regime, then this configuration will typically lead to images where the current is zero except for ring-shaped regions where one of the quantized energy levels of the dot is in resonance with the chemical potential of source and drain, and the dot conductance is enhanced \cite{Woodside:2002,Pioda:2004,Kicin:2005,Fallahi:2005}. In Fig. \ref{fig:potential}(a) we show such a measurement. 

In this configuration the tip acts as a plunger gate and we will argue below that the ring-shaped regions of high current are equipotential lines of the tip potential as one would expect. In principle, this technique can be used to determine the tip potential, but two challenges usually prevent a full quantitative analysis. On the one hand, one cannot measure the potential in between two rings because here the dot is simply not a sensitive detector. On the other hand, the energy separation between the quantized levels of the dot is not uniform, especially for small few-electron dots, and so one does not know what the potential is at a given equipotential line. In Ref. \cite{Gildemeister2:2007} we have used an unconventional Coulomb diamond measurement to determine the tip potential, but this technique is only practicable in one dimension.

Here we have used a quantum dot in the Coulomb blockade regime as a sensitive potentiometer to quantitatively measure the  potential of a metallic SFM tip with high resolution over an area of up to $2\times2\ \mu$m$^2$.  The tip was scanned at a constant height of about 200~nm over the sample surface and we used a feedback mechanism to apply a voltage to a plunger gate such that one of the quantized energy levels of the dot would always stay in resonance with the chemical potential of the source and drain leads. The voltage on the plunger gate corresponds to the tip potential. With this technique we could ensure that the dot was a sensitive detector for every tip position and that we used only one quantum state for detection.    

Additionally, we measured the lever arm of the tip as a function of position by applying an ac voltage to the tip and measuring how strongly the feedback of the plunger gate reacted to this. The lever arm helps to understand the origin of the tip potential, its behavior as a function of the voltage applied to the tip, and the contact potential difference between tip and sample.

The experiments were carried out during a single cooldown in a dilution refrigerator cooled SFM \cite{Gildemeister1:2007} with an electrochemically etched PtIr tip. The electronic temperature of the sample was about 190~mK as determined from the width of conductance resonances in the Coulomb blockade regime.

\begin{figure}[ht]
\includegraphics[width=8.2cm]{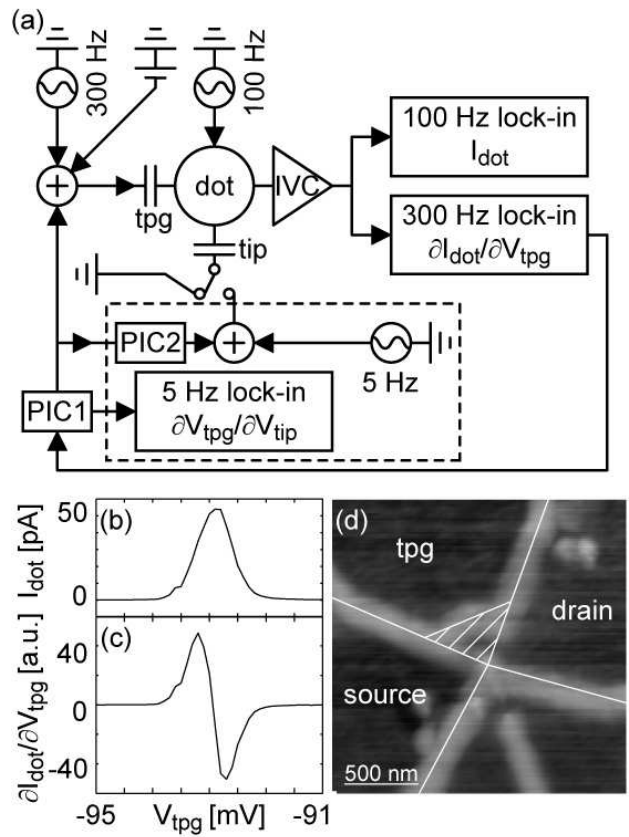}
\caption{\label{fig:technique} (a) Schematic circuit diagram of the experimental setup. (b) The current through the quantum dot $\Idot$ as a function of the voltage $\Vtpg$ applied to the top plunger gate (tpg) and (c) the measured derivative of $\Idot$ with respect to $\Vtpg$. (d) Topography scan of the sample recorded at base temperature. Oxide lines protruding from the sample surface define the structure in the 2DEG underneath. The surface is covered with a thin Ti film which is again patterned by oxidation along the thin white lines. In the center of the image we can see the quantum dot which is tunnel-coupled to source and drain leads. In the triangular area above the dot the Ti film is fully oxidized. The upper segment of the Ti film is used as the top plunger gate. Below the quantum dot a quantum point contact can be seen. It may be used as a charge detector, but it is not discussed here. }
\end{figure}

\subsection{Sample Details}
The sample was prepared on a GaAs/AlGaAs heterostructure with a two-dimensional electron gas (2DEG) residing 34~nm below the surface. The electron density of the wafer was $n = 5.3\times10^{15}$~m$^{-2}$ and the mobility was $\mu=30$~m$^2/$Vs at 4.2~K. 

Figure \ref{fig:technique}(d) shows a  topography scan of the sample obtained at the base-temperature of the dilution refrigerator. At room temperature a quantum dot was patterned by local anodic oxidation of the GaAs surface \cite{Held:1997}. Then a thin Ti film was evaporated on the surface and this film was again patterned by local anodic oxidation to form top gates \cite{Sigrist:2004}. The Ti film in the area directly above the quantum dot was fully oxidized. It was through this ``window" that the tip could interact capacitively with the dot. 

The top gates bring two advantages compared to in-plane gates. First, we achieve more tunability of the quantum dot. Second and more important, impurities that can act as charge traps \cite{Pioda:2007,Gildemeister2:2007} are screened from the tip potential. This effectively prevents parametric rearrangements of charges in the sample, an effect that can impair data quality and fosters misinterpretations. Of four top gates we used three: Two to tune the tunnel-coupling to the source and drain leads of the dot and one as the top plunger gate (tpg).

\section{Tip-Induced Potential}

\subsection{How to measure the tip potential}

Figure \ref{fig:technique}(a) gives an overview of the measurement setup. For measuring the tip potential the components within the dashed box are not in use and the tip is grounded. 

Tunnel barriers couple the dot to source (top) and drain (right). We apply a small ac bias of 20~$\mu$V at about 100~Hz between source and drain. The bias across the dot creates a current and we use a current-to-voltage converter (IVC) to measure it with a lock-in amplifier that demodulates at about $100$~Hz and thereby differentially measures the current $\Idot$ through the dot. This is a standard technique.

The dot is capacitively coupled to several gates of which we consider only the top plunger gate  and the tip of the microscope. While the tip is grounded, we apply the sum of three voltages to the tpg. First, we apply a dc voltage $\Vtpg$ that is needed to tune the dot to the Coulomb blockade regime. Second, we apply an ac voltage of 0.2~mV at about 300~Hz to measure the derivative $\partial \Idot / \partial \Vtpg$ of the current through the dot with respect to the voltage $\Vtpg$ applied to the tpg. Third, we apply a feedback voltage $\Vfb$ that we will discuss in more detail.

Figure \ref{fig:technique}(b) shows a typical trace of the current through the dot as a function of $\Vtpg$. We show one resonance. Figure \ref{fig:technique}(c) shows the simultaneously measured $\partial \Idot / \partial \Vtpg$ which exhibits a peak where $\Idot$ rises and a dip where $\Idot$ falls. 

We wish to set up a feedback that drives $\Vfb$ in such a way that the dot always remains at resonance, even when the tip moves. We achieve this by using  $\didv$ as the error signal. The working point is at the center of the resonance, where $\Idot$ is maximal and $\didv$ is zero.  If $\didv$ becomes positive (negative) then $\Vfb$ is too low (high) and should be increased (decreased). In order to obtain a stable feedback, the $\didv$ signal needs to be appropriately filtered. This is done by the first proportional and integral controller (PIC1) \footnote{We assume a PIC with a time-domain response $\Vfb (t) = P \ \didv (t) + I \int_{-\infty}^t dt' \ \didv (t')$, where $P$ is the proportional coefficient and $I$ the integral coefficient. We define the time constant $T$ of the PIC as $T=P/I$.}.

To achieve a stable feedback the time constant $T$ of PIC1 should be set to $T=\tau$, where $\tau$ is the time constant of the lock-in that measures $\didv$. The bandwidth $b$ of the feedback is $$b=-\frac{ P}{2 \pi T}\frac{\partial^2 \Idot }{ \partial \Vtpg^2},$$ where $P$ is the proportional coefficient of PIC1. The noise on $\Vfb$ must not exceed the width of the resonance because otherwise the feedback would lose the resonance and collapse. In order to achieve this, one has to choose a sufficiently small $P$. The resulting bandwidth is $b \approx 80$~Hz. 

When we scan the tip over the sample then it acts as a plunger gate and shifts the energy levels of the dot. With the feedback turned on, $\Vfb$ will compensate the shift and the dot stays at resonance. Therefore the output $\Vfb$ of the feedback tells us what the tip potential is. With Coulomb diamond measurements we have determined the lever arm of the tpg to be $\atpg=9.5\%$ in absence of the tip. We assume that $\atpg$ changes very little as the tip is scanned over the dot, as will be justified below. Then the tip potential is
$$
\Phi  = - \atpg \Vfb 
$$
and the potential energy of an electron in the field of the tip is $-e \Phi  =  \atpg e \Vfb $, where $e$ is the elementary charge.

\subsection{Results}

\begin{figure}[tb]
\includegraphics[width=8.2cm]{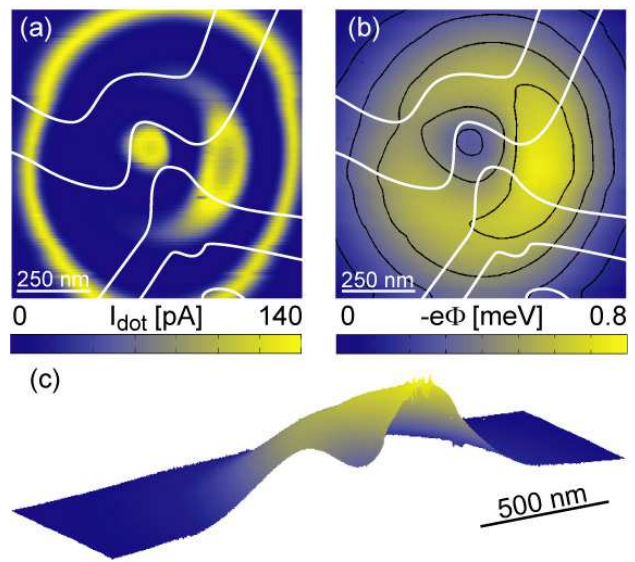}
\caption{\label{fig:potential} (Color online) (a) Conventional scanning gate image of the current through a quantum dot. The grounded tip is moved over the dot at a constant height of about 200~nm above the sample surface. The oxide lines that define the quantum dot in the center are outlined in white. (b) The tip potential measured with the quantum dot. Equipotential contours are plotted as black lines. (c) Sectional view of the tip potential with a larger scan range. }
\end{figure}

In Fig. \ref{fig:potential}(b) we show the tip potential  that was measured with the feedback turned on. We have multiplied the potential with $-e$ to show the more intuitive potential energy. On the large scale we see a roughly circularly symmetric repulsive potential peak which is about 700~nm wide and about 1~meV high. Close to the center we see a smaller attractive potential dip superimposed on the large peak. The dip is also circularly symmetric, about 250~nm wide, and around 0.5~meV deep. In Fig. \ref{fig:potential}(c) we show a sectional view of the tip potential from another measurement with a larger scan-range. This measurement shows that further away from the center the tip potential falls off and becomes almost flat. The additional measurement also demonstrates the reproducibility of the data.

Because the bandwidth of the feedback is fairly small, the tip had to be scanned at a slow pace of 2~nm/s and measuring the tip potential took 13 hours. This shows that the feedback and the dot can be very stable for a long time.

In Fig. \ref{fig:potential}(a) we show a scanning gate measurement of the quantum dot where we simply recorded the current through the dot as a function of tip position. The feedback was turned off and the grounded tip was scanned over the sample surface at a height of about 200~nm.

When we compare the tip potential in Fig. \ref{fig:potential}(b)  with the current image of Fig. \ref{fig:potential}(a) then it is evident that there is high current along equipotential lines of the tip potential. The interpretation of a current image becomes clearer when the tip potential is known. For example, it is now possible to associate the very small ring in the middle and the large ring with the same charge state of the quantum dot, whereas the crescent-shaped arc corresponds to a state with one electron less on the dot.

Clearly, this potential does not fulfill the requirements for probability density mapping defined above. The potential is wider than the dot and, as we can see in the current image, it exceeds the charging energy of the dot. Also, it has both an attractive and a repulsive part which makes it complicated. In view of future experiments it is important to understand the origin of the tip potential and possibilities to improve it.

\section{Lever Arm of the Tip}

\begin{figure}[t]
\includegraphics[width=8.2cm]{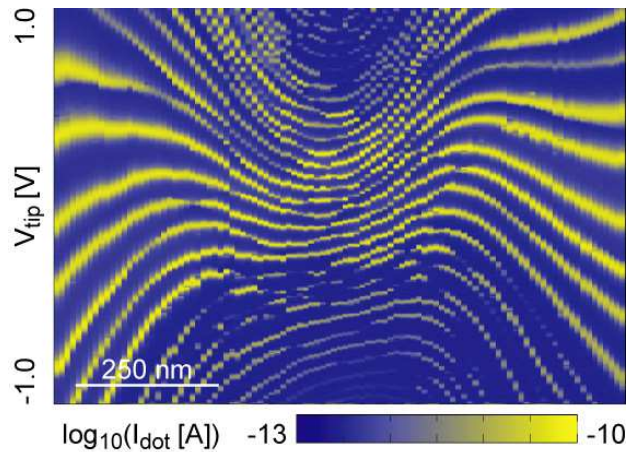}
\caption{\label{fig:vdep} (Color online) Plot of the current through the dot with a logarithmic color scale. The tip was moved stepwise along a horizontal line across the center of the dot. For every point the voltage applied to the tip was swept. The plot shows the current as a function of tip position and tip voltage. }
\end{figure}

Naively one would expect the tip potential to be the product of the spatially variable lever arm of the tip and the voltage applied to the tip, where the latter would have to be corrected by the contact potential difference $\Vcpd$ between the sample and the tip. If this were the case then one could null the tip potential by compensating $\Vcpd$.

However, it is an empirical observation \cite{Pioda:2004,Finkelstein:2000} that the tip potential cannot be nulled for all tip positions by applying an appropriate voltage to the tip. To illustrate this we show in Fig. \ref{fig:vdep} a measurement where the current through the dot was recorded as a function of the tip position and the voltage $\Vtip$ applied to the tip. The tip was moved stepwise along a line in $x$-direction across the center of the dot and for every step of the tip $\Vtip$ was swept from $-1$~V to $+1$~V. About 30 wavy lines of high dot current can be seen. Each line corresponds to one quantized state of the dot and follows an equipotential line of the tip. Applying sufficiently large positive or negative voltages can make the potential either attractive or repulsive, which leads to convex and concave lines at the top and bottom in Fig. \ref{fig:vdep}. However, there are no straight horizontal lines in between. This means that one cannot find a $\Vtip$ at which the tip does not influence the quantum dot. 

We conclude that the tip potential consists of two parts. The first part, $\Phid$, depends on $\Vtip$, the second part, $\Phiin$, is independent of $\Vtip$. For the tip potential $\Phi$ and the tip's lever arm $\atip$ we write
\eq
\Phi(\Vtip,x,y)=\underbrace{\atip(x,y) (\Vtip-\Vcpd)}_{\Phid(\Vtip,x,y)} + \Phiin(x,y),
\label{eq:phi}
\qe
where we include that $\Vtip$ needs to be taken with respect to a constant contact potential difference. We assume that the tip is at position $(x,y)$ and at a constant height $z$ over the surface.

We have already measured $\Phi (x,y)$ and we will now turn to measuring the spatial dependence of the tip's lever arm $\atip$. A measurement of $\atip(x,y)$ together with an assumption about $\Vcpd$ will allow us to separate the two parts that contribute to the tip potential and to analyze them in more detail.

\subsection{How to measure the tip's lever arm}

In order to measure the tip's lever arm we used essentially the same setup that was described before for measuring the tip potential. The main difference is that now the tip is no longer grounded, but rather an ac voltage of 1~mV with a frequency of 5~Hz is applied to the tip. In the schematic of Fig. \ref{fig:technique}(a) the switch is turned and the dashed box is now active. The 5~Hz lock-in measures how much the feedback has to change $\Vfb$ in order to compensate the ac voltage on the tip. Thereby it measures 
$$
\frac{\partial \Vtpg }{ \partial \Vtip} = \frac{\atip}{ \atpg},
$$
the lever arm of the tip relative to the lever arm of the top plunger gate.

We used a second feedback loop  to make the measurement more stable by minimizing the effect of the tip on the dot. A second proportional and integral controller (PIC2), not used in the measurement of the tip potential, was now used with a bandwidth well below 1~Hz. As the tip moved slowly over the sample, its potential was compensated by the voltage on the tpg. In order to disturb the dot as little as possible, the second feedback applied a quasi-dc voltage to the tip so that the compensation of the tip potential was now done by the tip itself. Only the 5~Hz ac modulation of the tip voltage remained to be compensated by the tpg because it was above the bandwidth of PIC2. With this configuration we could measure the relative lever arm of the tip without shifting the energy of the quantized state of the dot.

\subsection{Results}

\begin{figure}[t]
\includegraphics[width=8.2cm]{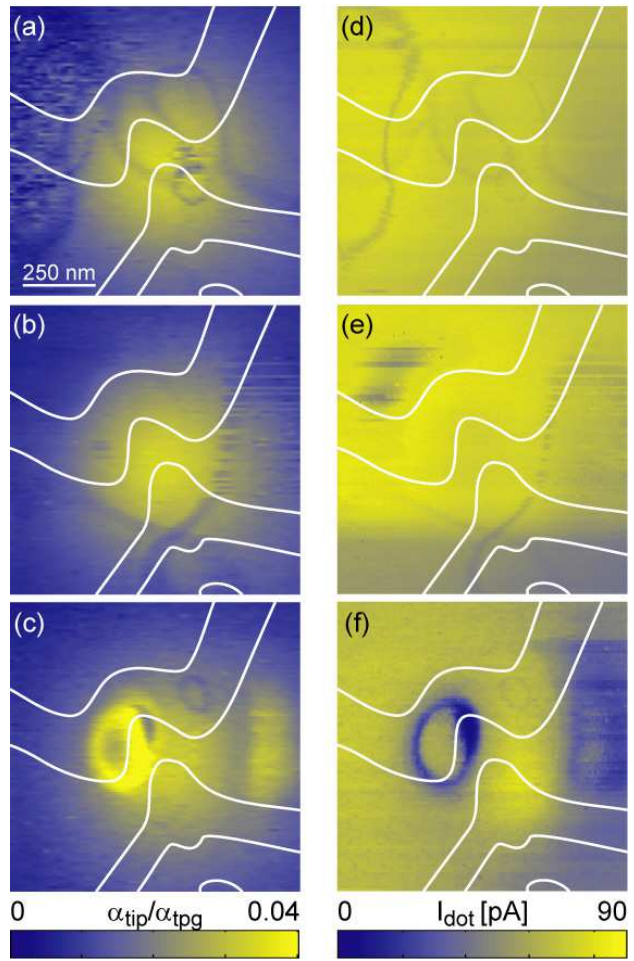}
\caption{\label{fig:leveram} (Color online) (a-c) Measurements of the lever arm of the tip relative to the lever arm of the top plunger gate for three different quantum states of the dot. (d-f) Simultaneously recorded current through the dot.}
\end{figure}

In Figs. \ref{fig:leveram}(a,b,c) we show the tip's relative lever arm measured with three different quantum states of the dot. Generally, the lever arm is highest when the tip is directly over the dot and it falls off when the tip is moved away from the dot. 

An average of the three measurements is shown in Fig. \ref{fig:reconstruct}(b). The average lever arm can be described as a rather symmetric Lorentzian peak, centered over the dot, about 700~nm wide and with a maximum of $\atip/\atpg \approx 4\%$. We know that in absence of the tip $\atpg \approx 9.5\%$. Because of screening the tip will slightly reduce the lever arm of the tpg. Nevertheless,  since $\atip/\atpg \ll 1$, it is a good approximation to  regard the tpg lever arm as constant. Under this assumption we find that $\atip \approx 0.4 \%$ at maximum. This value is smaller than in previous reports \cite{Kicin:2005,Gildemeister2:2007} because the tip is screened by the top gates. We will see below how the measurement of $\atip(x,y)$ can be used to model the tip potential. 

While the general spatial dependence of the lever arm remains similar when measured with different quantum states of the dot, we see a lot of fine structure in the measurements of $\atip/\atpg$ that is different for each state. 
All three measurements show wavy open lines of low $\atip/\atpg$ and Figs. \ref{fig:leveram}(a) and \ref{fig:leveram}(c) show small circles of low $\atip/\atpg$ at different positions.
At present we can only speculate about the origin of this fine structure.  

For comparison we show the current through the dot $\Idot$ that was simultaneously measured with $\atip/\atpg$ in Figs. \ref{fig:leveram}(d-f). We find that most lines of low $\atip/\atpg$ coincide with lines of low $\Idot$. A low dot current decreases the bandwidth of our feedback and possibly the low dot current could be the reason for a low measured lever arm. However, two arguments speak against this. First, in Figs. \ref{fig:leveram}(c,f) we find an example, where, in the large oval on the left, $\atip/\atpg$ is particularly high while the current is particularly low. Second, we could reproduce the fine structure of Fig. \ref{fig:leveram}(a) even when the tip voltage was modulated at 1~Hz instead of the otherwise used 5~Hz. Even if it is the low dot current which generally causes the low measured lever arm, then the question remains, why the dot current would decrease along these particular lines.

Obviously, the quantum states used in the measurements differ in the shape of their wave function. Therefore it would be possible that the properties of the wave function could leave some kind of ``quantum fingerprint" in the measurement of $\atip/\atpg$. However, the width of the tip potential and the length scales of the fine structure exclude any interpretation of the data in the sense of probability density mapping.

\section{Analysis of the Tip Potential}

\begin{figure}[ht]
\includegraphics[width=8.2cm]{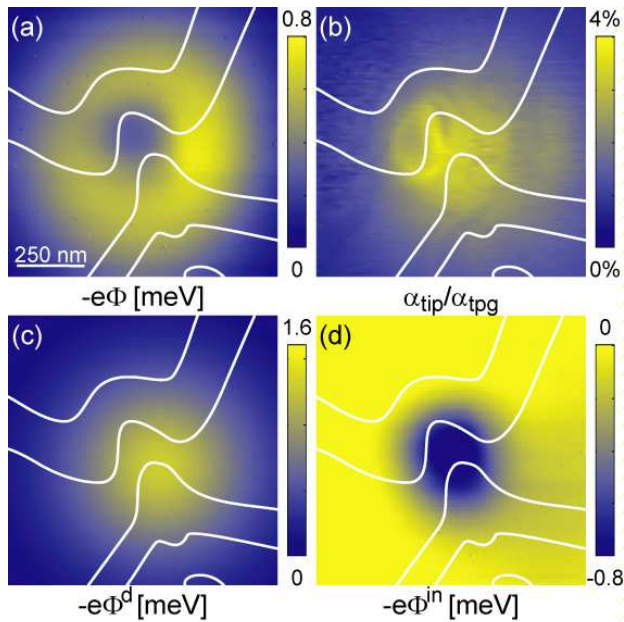}
\caption{\label{fig:reconstruct} (Color online) (a) Tip-induced potential from Fig. \ref{fig:potential}(b) for comparison. (b) Average over the measurements of the lever arm of the tip in Figs. \ref{fig:leveram}(a-c). (c) Fit to the tip voltage dependent part of the potential $\Phid$. (d) Tip voltage independent part of the potential $\Phiin   $.}
\end{figure}

We now have the quantitatively measured tip potential $\Phi$ as well as the tip's lever arm $\atip$ at hand. We can use these quantities to determine the two parts of the tip potential $\Phid$ and $\Phiin$. 

When we measured $\Phi$ the tip was grounded and Eq. (\ref{eq:phi}) tells us that in this case $\Phid(0,x,y)=-\atip(x,y) \Vcpd$. In order not to overestimate the fine structure found in the measurements of $\atip(x,y)$ we have fitted a Lorentzian curve to the average of the three measurements in Figs. \ref{fig:leveram}(a-c) and used this fit to calculate $\Phid(x,y)$. For the contact potential difference we used $\Vcpd = 0.5$~V for two reasons. On the one hand, this value was obtained from Kelvin probe measurements in an experiment with identical tip and sample materials \cite{Vancura:2003}. On the other hand, this choice of $\Vcpd$ leads to a particularly simple spatial dependence of $\Phiin(x,y)$.

In Fig. \ref{fig:reconstruct}(a) we again show the tip potential $\Phi(x,y)$ of Fig. \ref{fig:potential}(b) for comparison. In Fig. \ref{fig:reconstruct}(b) we show the tip's relative lever arm, averaged over the three measurements shown in Figs. \ref{fig:leveram}(a-c). Figure \ref{fig:reconstruct}(c) shows the Lorentzian fit to the average $\atip/\atpg$, multiplied with  $\atpg$ and $\Vcpd$. This is $\Phid$ for the case of a grounded tip, 
$$
\Phid(0,x,y)=-\frac{\atip(x,y)}{\atpg} \atpg \Vcpd.
$$

We can now calculate $$\Phiin(x,y)=\Phi(0,x,y) - \Phid(0,x,y)$$ which is shown in Figure \ref{fig:reconstruct}(d). In contrast to $\Phid(x,y)$ we see that $\Phiin(x,y)$ is attractive. It can be described as a roughly circularly symmetric Lorentzian dip that is about 250 nm wide and 0.8 meV deep.

Our experience shows that the tip potential can change if the tip is used for topography scans. From other measurements \cite{Gildemeister4:2007} it is also known that a high-field treatment of the tip \cite{Chen:1993}, i.e., suddenly applying relatively large voltages between the tip and a metallic sample surface, can modify the tip potential.

An obvious explanation for this behavior would be an electrically charged dielectric particle that clings to the metallic tip. Such particles could be moved around during topography scans and would create a potential that is independent of the voltage applied to the tip. In Ref. \cite{Finkelstein:2000} the authors suggest that GaAs debris from the surface could get picked up by the tip.

In Fig. \ref{fig:technique}(c) we show a topography scan of the structure which was recorded immediately after the other measurements presented here. The resolution of the topography scan and the absence of a topographical double tip suggest that the particle has to be either small compared to the tip or set back from the tip far enough so that it does not touch the surface. 

We can estimate the charge of the particle that would be necessary to create the observed potential. While an exact description of the electrostatics would be rather involved, we can make an order of magnitude estimate. We assume that the particle is $d_1=200$~nm above the sample surface and that the 2DEG is $d_2=34$~nm below the surface of the Ga[Al]As with its high dielectric constant $\epsilon=12.8$. Then the charge $q$ necessary to create a potential of $\Phiin = 0.8$~mV would be \cite{Jackson:1974}
$$
q \approx \frac{4 \pi \epsilon_0 (1+\epsilon)}{2} (d_1+d_2) \Phiin  \approx e.
$$  
We neglect all screening effects of the tip and the gates and therefore the real charge has to be higher. However, we see that a few elementary charges are able to create a significant change of the tip potential that cannot be compensated by applying a voltage to the tip.

While our quantitative analysis focuses on this particular cooldown and tip configuration we note that we have also observed tip voltage independent potentials with different tips on different quantum dots, in different cooldown cycles, and even in completely different SFM setups. Possibly, charged particles on the tip also account for the tip potentials observed in Refs. \cite{Woodside:2002, Pioda:2004, Pioda:2007}.

An evident measure to improve the tip potential is to move the tip closer to the surface because $\Phid$ would  presumably  be much narrower then. However, this will make the potential of a charged particle on the tip rather stronger. Unfortunately we could not measure the tip potential when the tip was closer to the surface because we could not achieve a stable feedback for the potential measurement then. To substantially improve the tip potential it would probably be necessary to create an absolutely clean metallic tip, avoid topography scans, and to move the tip very close to the sample surface.

\section{Conclusion}

We have presented an experimental technique that can be used to measure quantitatively  the spatial dependence of the potential induced in a quantum dot by the tip of a scanning force microscope. Furthermore, the technique allows one to quantitatively measure the spatial dependence of the lever arm of the tip. The feedback mechanism used for the measurements is able to minimize the tip potential and makes measurements in a least-invasive regime possible.

The tip potential generally can have two components, one that depends on the voltage applied to the tip and one that is independent of this voltage. In our measurements the tip voltage dependent part could be well described as the product of the tip's lever arm and the difference between the voltage applied to the tip and the contact potential difference between tip and sample. For a grounded tip it was a repulsive Lorentzian-shaped 700~nm wide and 1.6~meV high potential. The tip voltage independent potential was attractive, Lorentzian-shaped, about 250~nm wide, and 0.8~meV deep. It could be caused by a charged particle on the tip.

The measurements of the tip's lever arm revealed fine structure both in the lever arm and the dot current that was different for the three quantum states we measured. We speculate that this fine structure may be characteristic for a quantum state. 

The potential of the tip investigated here does not fulfill the requirements for probing the probability density of quantum states. Sharper and possibly cleaner tips are needed for such experiments. For the interpretation of results from any scanning gate experiment one should bear in mind that the tip potential could have an unexpected shape.

\begin{acknowledgments}
We would like to thank the team of Nanonis GmbH for their continual technical support. Financial support from ETH Z{\"u}rich is gratefully acknowledged.
\end{acknowledgments}



\end{document}